\documentclass[12pt,thmsa]{article}
\usepackage{amssymb}

\begin{document}

\title{A simple and universal setup of quasimomocolor gamma ray source}
\author{H. Lin, C. P. Liu, C .Wang. \\
{State Key Laboratory of High Field Laser Physics,}\\
Information Technology Research Center of Space Laser,\\
{Shanghai Institute of Optics and Fine Mechanics,}\\
{P. O. Box 800-211, Shanghai 201800, China.}\\
linhai@siom.ac.cn}
\maketitle

\begin{abstract}
Strict classic 3-D dynamics theory reveals that arbitrarily high center
frequency light source can be achieved by flexible application of
not-too-strong static electric field and static magnetic field. The
magnitudes of the fields are not required to be high.
\end{abstract}

Even though significant application value has promoted intensive
investigation on short-wavelength\ light source (or X-ray or higher energy
photons source) for decades [1-14], there are still considerable difficulty
in making a satisfactory achievement on this issue. Because $1$ $\mu m$ ($%
1nm $) wavelength radiation means the time cycle being $3.33fs$ ($3.33as$),
we therefore wish the time cycle of electron motion to be $3.33fs$ ($3.33as$%
). Clearly, the difficulty is just that so rapid electron oscillation, which
can be finished hundreds (or more) times per $fs$, is unavailable. This
limits the efficiency of generating X-ray. For example, in the high-order
harmonics generation (HHG), if the driving laser is of $1$ $\mu m$
wavelength, the driven atomic dipole moment oscillation is usually of a time
cycle $\symbol{126}3.33fs$. The HHG is only because the time shape of the
driven atomic dipole moment oscillation differs from $\sin \left( 2\pi \frac
t{3.33fs}\right) $greatly. This determines that low-order harmonics are
dominant. According to most HHG experimental results [7-14], low-order
harmonics components whose orders are tens can be warranted but components
at higher order hundreds are negligible. On the other hand, in
free-electron-laser (FEL) [2-5], the efficiency of generating radiations at
desired short wavelength is also far from satisfactory because sufficiently
rapid electron velocity oscillation is unavailable.

Clearly, if we can directly set up a sufficiently rapid electron velocity
oscillation, we will obtain an efficient monocolor short-wavelength\ light
source at desired short wavelength. If $1nm$ wavelength is desired, we
should set up an electron velocity oscillation whose time cycle is $\symbol{%
126}3.33as$. To drive so rapid oscillation usually demands, according to
current knowledge, electromagnetic field of very short wavelength, which is
unavailable because they are jsut what we are pursuing. Namely, we are
trying to get what we are pursuing by applying what we are pursuing.

We consider a simple configuration containing merely static electric field
(along $x$-direction) and static magnetic field (along $z$-direction). When
an electron is input into this configuration, its behavior can be described
by dimensionless 3-D\ relativistic Newton equations (RNEs) 
\begin{equation}
d_s\left[ \Gamma d_sZ\right] =0,  
\end{equation}
\begin{equation}
d_s\left[ \Gamma d_sY\right] =W_Bd_sX  
\end{equation}
\begin{equation}
d_s\left[ \Gamma d_sX\right] =-W_B\left[ \eta +d_sY\right]  
\end{equation}
where 
\begin{equation}
\frac 1\Gamma =\sqrt{1-\left( d_sX\right) ^2-\left( d_sY\right) ^2-\left(
d_sZ\right) ^2}.  
\end{equation}
Moreover, $E_s$ and $B_s$ are constant-valued electric field and magnetic
one and meet $E_s=\eta cB_s$, $\lambda =c/\omega $ and $\omega $ are
reference wavelength and frequency, and $s=\omega t,$ $Z=\frac z\lambda ,$ $%
Y=\frac y\lambda ,$ $X=\frac x\lambda ,W_B=\frac{\omega _B}\omega ,$ $\omega
_B=\frac{eB_s}{m_e}$ is the cyclotron frequency,.

Eqs.(1-3) will lead to

\begin{equation}
d_sZ\equiv 0  
\end{equation}
\begin{eqnarray}
\Gamma d_sY-W_BX &=&const=C_y;   \\
\Gamma d_sX+W_B\left[ \eta s+Y\right] &=&const=C_x,  
\end{eqnarray}
where the values of these constants $const$ are determined from the initial
conditions $\left( X,Y,Z,d_sX,d_sY,d_sZ\right) |_{s=0}=\left( 0,0,0,\frac{C_x%
}{\sqrt{1+C_x^2+C_y^2}},\frac{C_y}{\sqrt{1+C_x^2+C_y^2}},0\right) $.

Eqs.(5-7) can yield an equation set of $d_sX$ and $d_sY$

\begin{eqnarray}
\left( d_sY\right) ^2 &=&\left[ C_y+W_BX\right] ^2*\left[ 1-\left(
d_sX\right) ^2-\left( d_sY\right) ^2\right]   \\
\left( d_sX\right) ^2 &=&\left[ C_x-W_B*\left( \eta s+Y\right) \right]
^2*\left[ 1-\left( d_sX\right) ^2-\left( d_sY\right) ^2\right]  
\end{eqnarray}
whose solution reads 
\begin{equation}
\left( d_sX\right) ^2=\frac{\left[ C_x-W_B*\left( \eta s+Y\right) \right] ^2%
}{\left[ 1+\left[ C_y+W_BX\right] ^2+\left[ C_x-W_B*\left( \eta s+Y\right)
\right] ^2\right] }  
\end{equation}
\newline
\begin{equation}
\left( d_sY\right) ^2=\frac{\left[ C_y+W_BX\right] ^2}{\left[ 1+\left[
C_y+W_BX\right] ^2+\left[ C_x-W_B*\left( \eta s+Y\right) \right] ^2\right] }.
\end{equation}
Note that the solutions (10,11) will cause $\Gamma =\sqrt{%
1+\left[ C_y+W_BX\right] ^2+\left[ C_x-W_B*\left( \eta s+Y\right) \right] ^2}
$ and, with the help of Eqs.(6,7), $d_s\Gamma =-W_B\eta *d_sX$ (i.e. $%
m_ec^2d_t\Gamma =eEd_tX$). If noting $\Gamma $ can be formally expressed as $%
\Gamma =\sqrt{1+C_y^2+C_x^2}-W_B\eta *X$, which agrees with Takeuchi's
theory [15], we can find that the electronic trajectory can be expressed as 
\begin{equation}
\left[ \sqrt{1+C_y^2+C_x^2}-W_B\eta *X\right] ^2=1+\left[ C_y+W_BX\right]
^2+\left[ C_x-W_B*\left( \eta s+Y\right) \right] ^2,  
\end{equation}
or 
\begin{equation}
\left( 1-\eta ^2\right) \left[ X+\frac{\left( \eta +\upsilon _{y0}\right) }{%
1-\eta ^2}\frac{\Gamma _0}{W_B}\right] ^2+\left[ \left( Y+\eta s\right)
-\upsilon _{x0}\frac{\Gamma _0}{W_B}\right] ^2=\frac{\left[ \left( \eta
+\upsilon _{y0}\right) ^2+\left( 1-\eta ^2\right) \upsilon _{x0}^2\right] }{%
1-\eta ^2}\left( \frac{\Gamma _0}{W_B}\right) ^2,  
\end{equation}
where $\Gamma _0=\sqrt{1+C_y^2+C_x^2}$, $\upsilon _{x0}=\frac{C_x}{\Gamma _0}
$ and $\upsilon _{y0}=\frac{C_y}{\Gamma _0}$.

There will be an elliptical trajectory for $\eta <1$ and a hyperbolic one
for $\eta >1$ [15,16]. The time for an electron travelling through an
elliptical trajectory can be exactly calculated by re-writing Eq.(10) as
[15] 
\begin{equation}
\pm ds=\frac{\frac 1{W_B}\Gamma _0-\eta *X}{\sqrt{aX^2+bX+c}}dX=\frac \eta {%
\sqrt{-a}}\frac{X_N-X}{\sqrt{\frac{b^2-4ac}{4a^2}-\left( X+\frac
b{2a}\right) ^2}}dX,  
\end{equation}
where $a=\left( \eta ^2-1\right) $, $b=-2\left[ \eta \Gamma _0+C_y\right]
\frac 1{W_B},$ $c=C_x^2\left( \frac 1{W_B}\right) ^2$ and $X_N=\frac 1\eta
\frac 1{W_B}\Gamma _0$. The equation can be written as a more general form

\begin{equation}
\pm ds=\sigma \frac{M-u}{\sqrt{1-u^2}}du  
\end{equation}
where $u=\frac{X+\frac b{2a}}{\sqrt{\frac{b^2-4ac}{4a^2}}}=\frac{X-\frac{%
X_R+X_L}2}{X_R-X_L}$, $X_L=\min (\frac{-b-\sqrt{b^2-4ac}}{2a},\frac{-b+\sqrt{%
b^2-4ac}}{2a})$ and $X_R=\max (\frac{-b-\sqrt{b^2-4ac}}{2a},\frac{-b+\sqrt{%
b^2-4ac}}{2a})$. In addition, $\sigma =\frac \eta {\sqrt{-a}}\sqrt{\frac{%
b^2-4ac}{4a^2}}$ and $M=\frac{X_N+\frac b{2a}}{\sqrt{\frac{b^2-4ac}{4a^2}}}=%
\frac{X_N-\frac{X_R+X_L}2}{X_R-X_L}$. It is easy to verify that for $\eta
^2-1<0$, there is $M=\frac{1+\eta \upsilon _{y0}}{\eta \sqrt{\left( \eta
+\upsilon _{y0}\right) ^2+\left( 1-\eta ^2\right) \upsilon _{x0}^2}}>1$.
Initially, $\left( X,Y\right) |_{s=0}=\left( 0,0\right) $ and hence $%
u_{st}=u|_{s=0}=\frac{0+\frac b{2a}}{\sqrt{\frac{b^2-4ac}{4a^2}}}=\frac{%
-\left( \eta +\upsilon _{y0}\right) }{\sqrt{\left( \eta +\upsilon
_{y0}\right) ^2+\left( 1-\eta ^2\right) \upsilon _{x0}^2}}$. From strict
solution 
\begin{equation}
\pm s\left( u\right) =\sigma *\left\{ M*\arcsin \left( u\right) +\sqrt{1-u^2}%
\right\} +const,  
\end{equation}
we can find the time for an electron travelling through an elliptical
trajectory to meet $s_{cycle}=\omega T_c=2*\left[ \sigma M\pi \right] $ and
hence a time cycle $T_c=\frac{\left( 1+\eta \upsilon _{y0}\right) \Gamma _0}{%
\left( \sqrt{1-\eta ^2}\right) ^3}\frac{2\pi }{\omega _B}.$ Namely, the
oscillation along the elliptical trajectory will have a circular frequency $%
\leqslant \omega _B$. Moreover, it is interest to note that $\left( \upsilon
_{x0},\upsilon _{y0}\right) =\left( 0,-\eta \right) $ will lead to $\frac{%
\left[ \left( \eta +\upsilon _{y0}\right) ^2+\left( 1-\eta ^2\right)
\upsilon _{x0}^2\right] }{1-\eta ^2}=0$ and hence a straight-line trajectory 
$\left( X\left( s\right) ,Y\left( s\right) \right) =\left( 0,-\eta s\right) $%
.

The motion on an ellipical trajectory is very inhomogeneous. The time for
finishing the $\eta X>0$ half might be very short while that for the $\eta
X<0$ half might be very long. We term two halves as fast-half and slow-half
respectively. If $\eta $ is fixed over whole space, a fast-half is always
linked with a slow-half and hence makes the time cycle for finishing the
whole trajectory being at considerable level.

For convenience, our discussion is based on the parameterized ellipse. For
the case $\left( \upsilon _{x0}=0,\upsilon _{y0}=-\eta -\Delta \right) $,
(where $\Delta $ is small-valued positive), the starting position $X=0$ is
the left extreme of the ellipse and hence corresponds to $u=-1$. The time
required for an acute-angled rotation from $u=-1$ to $u=-1+\xi $, (where $%
\xi $ is small-valued positive), will be $\sigma M*\left[ \arcsin \left(
-1+\xi \right) -\frac \pi 2\right] +\sigma \sqrt{2\xi -\xi ^2}$, which is $%
=0 $ if $\xi =0$.

It is interest to note that if there is $B=0$ at the region $u>-1+\xi $, the
electron will enter from $\left( E,B=\frac E{\eta c}\right) $-region into $%
\left( E,B=0\right) $-region with an initial velocity whose $x$-component is 
$\upsilon _{x1}\symbol{126}d_su|_{u=-1+\xi }=\frac 1\sigma \frac{\sqrt{%
1-\left( -1+\xi \right) ^2}}{M-\left( -1+\xi \right) }>0$ and $y$-component $%
\upsilon _{y1}$ is $\neq 0$. Then, the electron will enter into the $\left(
E,B=0\right) $-region a distance because of $\upsilon _{x1}>0$. After a time 
$T_{tr}=\frac{2\upsilon _{x1}}{E\sqrt{1-\upsilon _{x1}^2-\upsilon _{y1}^2}}$%
, the electron will return into the $\left( E,B=\frac E{\eta c}\right) $%
-region and the returning velocity will have a $x$-component $-\upsilon
_{x1} $. During this stage, the electron will move $\upsilon _{y1}*T_{tr}$
along the $y$-direction. Then, the motion in the $\left( E,B=\frac E{\eta
c}\right) $-region can be described by an acute-angled rotation along the
ellipse $u=-1+\xi \rightarrow u=-1$. Thus, a complete closed cycle along the 
$x$-direction is finished even though the motion along the $y$-direction is
not closed. Repeating this closed cycle will lead to an oscillation along
the $x$-direction.

Clearly, the time cycle of such an oscillation is $T_x=T_{tr}+2\sigma
M*\left[ \arcsin \left( -1+\xi \right) -\frac \pi 2\right] +2\sigma \sqrt{%
2\xi -\xi ^2}$. Under fixed values of $\Delta $, $E$ and $B$, the smaller $%
\xi $ is, the smaller $T_x$ is. There will be $T_x=0$ at $\xi =0$. In
principle, arbitrary value of $T_x<T_c$ can be achieved by choosing suitable
value of $\xi $. Namely, arbitrary high center frequency ($>\omega _B$)
oscillation can be achieved by choosing suitable value of $\xi $. Although
the time history of $x\left( t\right) $ might cause its Fourier spectrum
being of some spread, the center frequency will be $\frac 1{T_x}$.

This result implies a simple and universal method of setting up
quasi-monocolor light source at any desirable center wavelength. That is,
applying vertically static electric field $E=E_x$ and static magnetic field $%
B=B_z$ and on purpose letting a $B=0$ region existing and the ratio $\frac
E{cB}<1$, then injecting electron along the $y$-axis with a velocity
slightly above $|\frac E{cB}|$, and close to the boundary line between the $%
B=0$ region and the $B\neq 0$ region. As shown in Fig.1, adjusting the
distance $D=\xi *\sqrt{\frac{b^2-4ac}{4a^2}}$ can lead to a quasi-monocolor
oscillation source with any desired center frequency up to $\gamma $-ray
level.

In conclusion, we have described a simple and universal method of achieving
quasi-monocolor light source at any desirable center wavelength. The kernel
of this method is to utilize the motion of fast electron near the surface of
magnetic field. Here, fast electron means its velocity being $>\frac E{cB}$.
Actually, it is to adopt the fast-half of an ellipse and replace the
time-consuming slow-half with a faster orbit governed by $E$ only.

\section{Acknowledgment}

This work is supported by National Science Fund no 11374318.

Fig.1. The sketch of experimental setup.

\end{document}